\documentclass[twocolumn]{jpsj2} 
\title{Harmonic Content of Strain-induced Potential Modulation in Unidirectional Lateral Superlattices}

\author{Akira \textsc{Endo}\thanks{E-mail address: akrendo@issp.u-tokyo.ac.jp} and Yasuhiro \textsc{Iye}}

\inst{Institute for Solid State Physics, University of Tokyo, 5-1-5 Kashiwanoha, Kashiwa, Chiba 277-8581}

\abst{Detailed analysis of the commensurability oscillation (CO) has been performed on unidirectional lateral superlattices with periods ranging from $a$=92 to 184 nm. Fourier analysis reveals the second (and the third) harmonics along with the fundamental oscillation for $a$$\geq$138 nm (184 nm) at low-enough temperature, evincing the presence of corresponding harmonics in the profile of the potential modulation. The harmonics manifest themselves in CO with demagnified amplitude due to the low-pass filtering action of the thermal damping factor; with a suitable consideration of the damping effect, the harmonics of the modulation potential are found to have the amplitudes $V_2$ and $V_3$ up to roughly 30\% of that of the fundamental component $V_1$, despite the small ratio of the period $a$ to the depth $d$$\simeq$99 nm of the two-dimensional electron gas (2DEG) from the surface. The dependence of $V_n$ on $a$ indicates that the fundamental component originates at the surface, while the higher harmonics arise from the effect of the strain that penetrates down into subsurface. The manipulation of high harmonics thus provides a useful technique to introduce small length-scale modulation into high-mobility 2DEGs located deep inside the wafer.}

\kword{lateral superlattice, commensurability oscillation, harmonics, strain, stress, piezoelectric effect, two-dimensional electron gas, GaAs, AlGaAs}

\begin{document}
\maketitle

\section{Introduction} 
A high-mobility two-dimensional electron gas (2DEG) formed in a GaAs/AlGaAs-based heterostructure, often quoted also as modulation-doped field-effect transistor (MODFET), represents arguably the cleanest electron systems in solid state materials that have ever been made. The record mobility has reached a value as high as $\mu$=3,100 m$^2$/Vs \cite{Pfeiffer03}, which amounts to the electron mean-free path exceeding 0.1 mm. A MODFET, combined with modern nano-fabrication technologies, forms the basis of a wide range of experimental studies on the behavior of electrons under a designed environment \cite{BeenakkerR91}. In the majority of cases, the electrons in 2DEG are controlled by a patterned gate placed on the surface of the 2DEG wafer. Sizable efforts have been directed toward making the length scale thus introduced smaller, in pursuit of phenomena observable only when the artificial structure has a small-enough length scale \cite{Geisler04}. A serious drawback of MODFET in this respect is the fact that the 2DEG reside at the depth of tens or hundreds of nanometers from the surface; this is inevitable since the crux of attaining high mobility in a MODFET is to set back sources of electron-scattering, the ionized donor dopant (Si) being the most operative, as far away from the 2DEG channel as possible. The large distance of 2DEG from the surface renders it a formidable task to introduce a length scale smaller than the depth using nano-fabricated gates on the surface, since effects originating at the surface are generally expected to decay exponentially with the ratio of the depth $d$ to the lateral length scale $a_\mathrm{lateral}$ as $\propto\exp(-2\pi d/a_\mathrm{lateral})$.

A unidirectional lateral superlattice (ULSL) typifies the combined system of nano-fabricated gates and MODFET, where one-dimensionally modulated potential is introduced to 2DEG by a grating placed on the surface. ULSLs offer a unique opportunity to evaluate, through the amplitudes of commensurability oscillation (CO) appearing in the low-field magnetoresistance \cite{Weiss89}, the magnitude of the potential modulation seen by the electrons \cite{Beenakker89,Peeters92,Endo00e}. In the present paper, we investigate, by examining the CO in ULSLs, the harmonic content of the modulated potential introduced by elastic strain arising from differential contraction between the gate material and GaAs \cite{Yagi93,Ye95,Davies94,Kato97,Larkin97,Skuras97,Suzuki01}. The period $a$ of ULSL samples are chosen to be relatively small so as to be close to the depth $d$$\simeq$99 nm. Although the fundamental component resulting from the modulation period $a$ equal to that of the grating dominates the CO, second and third harmonics corresponding to periods $a/2$ or $a/3$ are also detected, more clearly when the temperature is lowered. We discuss the importance of the thermal damping on the detectability in CO of the smaller period modulation. Modulation amplitudes $V_n$ ($n$=1, 2, 3) for the fundamental component and harmonics are explored as a function of $a$, which hints at the origin of each component. The harmonics turn out to be more persistent than simple exponential decay mentioned above. Noticeably, $a/2$ and $a/3$ are significantly smaller than $d$. Therefore exploitation of high harmonics proves to be a powerful tool for introducing length scale smaller than the depth into the high-quality deeply residing 2DEGs.

\section{Experimental}
Five ULSL samples with differing periods ($a$=92, 115, 138, 161, and 184 nm) were prepared from the same Al$_{0.33}$Ga$_{0.67}$As/GaAs single-heterostructure 2DEG (MODFET) wafer with the mobility and electron density $\mu$$\simeq$70 m$^2$/Vs and $n_\mathrm{e}$=2.0$\times$10$^{15}$ m$^{-2}$ at 4.2 K, respectively. The structure of the wafer was (from the front surface) 10 nm GaAs cap layer, 40 nm Si-doped ($N_\mathrm{Si}$=2$\times$10$^{24}$ m$^{-3}$) Al$_{0.33}$Ga$_{0.67}$As layer, 40 nm undoped Al$_{0.33}$Ga$_{0.67}$As spacer layer, and 1 $\mu$m GaAs layer with 2DEG channel residing near the interface to the upper layer. The electron density, hence the mobility, was varied by illumination with infrared light-emitting diode (LED) when necessary. The potential modulation was introduced by placing a grating of high-resolution negative electron-beam (EB) resist (calixarene) \cite{Fujita96} on the surface of a 2DEG wafer patterned as a Hall bar by wet etching. The grating was oriented perpendicular to the direction of current so that the current flows across the modulation. The direction of the current was chosen to be a $\langle$110$\rangle$ direction. On cooling down the sample to cryogenic temperatures for measurement, elastic strain results from differential contraction between the resist and GaAs, which couples to 2DEG by deformation potential \cite{Davies94} or piezoelectric effect \cite{Larkin97} to introduce modulation, the latter mechanism being maximized and dominant for our choice of the crystallographic direction \cite{Skuras97}. The depth $d$$\simeq$99 nm of the 2DEG from the surface was evaluated as the sum of the distance between the surface and heterointerface, 90 nm, and the average distance of the 2DEG wave function from the interface, the latter being assessed as 8.7 nm (at $n_\mathrm{e}$=2.0$\times$10$^{15}$ m$^{-2}$) by a numerical self-consistent calculation including exchange and correlation effect \cite{Stern84,Endo05m}. Note that the values of $a$ are kept smaller than twice the $d$ in all of our ULSL samples. Magnetoresistance measurements were carried out at either 1.4 or 4.2 K (separate cryostats were used for each temperature), employing a standard low-frequency ac lock-in technique. We assume throughout the paper that the state of elastic strain hence the modulation amplitude do not differ between the two temperatures.

\section{Results}
\begin{figure}[tb]
\includegraphics[bbllx=20,bblly=20,bburx=420,bbury=820,width=8.5cm]{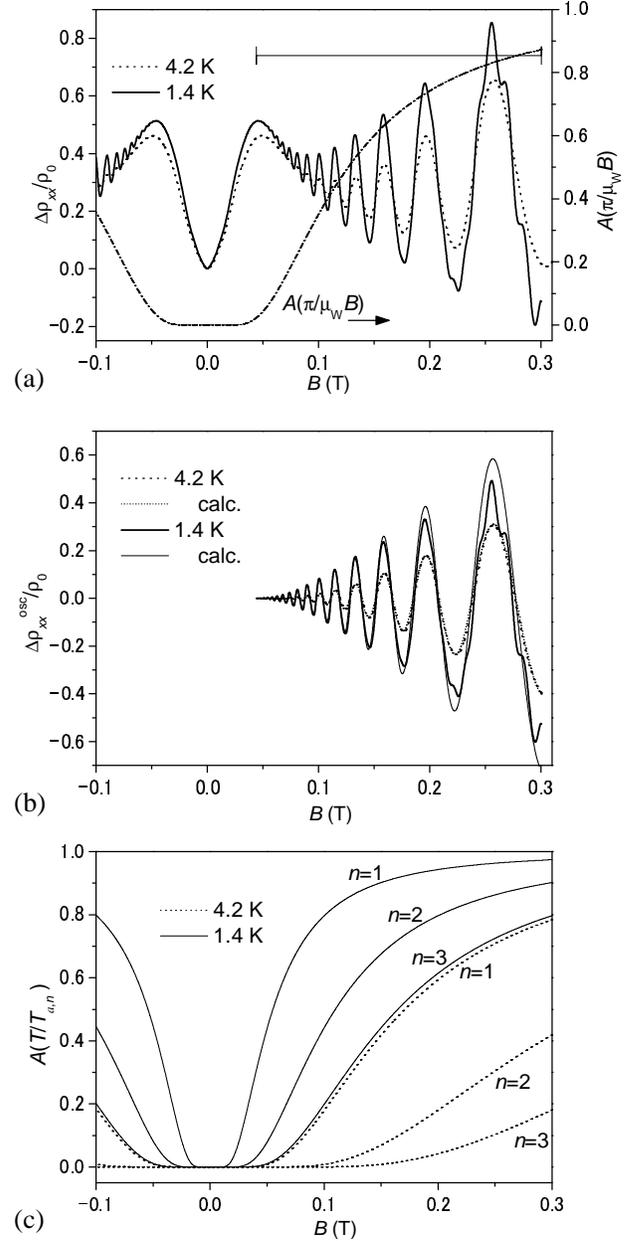}
\caption{(a) Magnetoresistance (left axis) at 1.4 K (solid trace) and 4.2 K (dotted trace) for ULSL with $a$=184 nm and $n_\mathrm{e}$=2.3$\times$10$^{15}$ m$^{-2}$. The horizontal marker indicates the window to be used for the Fourier analysis. The damping factor $A(\pi/\mu_\mathrm{W}B)$ (right axis) due to scattering is plotted by dash-dotted line for $\mu_\mathrm{W}$=11 m$^2$/Vs obtained by fitting to the 1.4 K trace. (b) Oscillatory part of magnetoresistance obtained by subtracting a slowly-varying background. Calculated traces using eq.\ (\ref{magres1}) are also plotted by thin solid (1.4 K) and dotted (4.2 K) lines. (c) Thermal damping factors $A(T/T_{a,n})$ at 1.4 K (solid lines) and 4.2 K (dotted lines) for $n$=1, 2, and 3. See eqs.\ (\ref{magres}) and (\ref{chrtmp}) \cite{separate}.}
\label{rawtherm}
\end{figure}

Figure \ref{rawtherm}(a) shows typical magnetoresistance traces at 1.4 and 4.2 K, exhibiting positive magnetoresistance (PMR) \cite{Beton90P} emanating from $B$=0 and the commensurability oscillation (CO), the latter being the subject of main interest in the present paper. In a previous paper \cite{Endo00e}, the present authors have shown that CO for sinusoidal potential modulation $V(x)$=$V_1\cos(2\pi x/a)$ is well described by,
\begin{equation}
\frac{\Delta\rho_{xx}^\mathrm{osc}}{\rho_0}=\gamma A\left(\frac{\pi}{\mu_\mathrm{W}B}\right) A\left(\frac{T}{T_{a,1}}\right) |B| \frac{V_1^2}{a}\sin\left(2\pi\frac{2R_\mathrm{c} }{a}\right),
\label{magres1}
\end{equation}
where $R_\mathrm{c}$=$\hbar k_\mathrm{F}/e|B|$ represents the cyclotron radius with $k_\mathrm{F}$=$\sqrt{2\pi n_\mathrm{e}}$ the Fermi wave number, and $\gamma$ a constant determined by sample parameters $\mu$ and $n_\mathrm{e}$ as
\begin{equation}
\gamma=\frac{1}{2\sqrt{2\pi}}\left(\frac{h}{e}\right)^{-1}\left(\frac{e\hbar}{2m^*}\right)^{-2}\frac{\mu^2}{n_\mathrm{e} ^{3/2}}.
\label{cnst}
\end{equation}
The factors $A(\pi/\mu_\mathrm{W}B)$ and $A(T/T_{a,1})$ specified using the function $A(x)$$\equiv$$x/\sinh(x)$ express damping due to scattering and temperature, respectively,  of the oscillation amplitude, which would otherwise decrease simply in proportion to the decreasing magnetic field. The parameter $\mu_\mathrm{W}$ is a measure of the scattering that diverts electrons away from the cyclotron orbit and have been shown \cite{Endo00e} to be essentially identical with the quantum mobility $\mu_\mathrm{Q}$ deduced from the analysis of the damping of the Shubnikov-de Haas (SdH) oscillation \cite{Coleridge91}. The characteristic temperature $T_{a,1}$ for the thermal damping factor is given by $k_\mathrm{B}T_{a,1}$=$(1/2\pi^2)(ak_\mathrm{F}/2)\hbar\omega_\mathrm{c}$ with $\omega_\mathrm{c}=eB/m^*$ the cyclotron angular frequency. The temperature differs from its counterpart in the SdH oscillation $T_\mathrm{c}$ only by the factor $ak_\mathrm{F}/2$, and therefore the sensitivity of CO to temperature approaches that of SdH for small $a$. The dependence of $A(\pi/\mu_\mathrm{W}B)$ and $A(T/T_{a,1})$ on $B$ is plotted in Figs.\ \ref{rawtherm}(a) and \ref{rawtherm}(c), respectively. The figures clearly demonstrate that the damping factors are indispensable for the understanding the envelope of CO; particularly, the thermal damping factor $A(T/T_{a,1})$ \cite{Beton90T,Peeters92}, although often neglected in attempts to explain CO amplitudes \cite{Boggild95,Paltiel97,Long99}, takes on greater importance as the period $a$ becomes smaller.

As illustrated in Fig.\ \ref{rawtherm}(b), the oscillatory part of the magnetoresistance, obtained by subtracting a slowly-varying background following the prescription delineated in ref.\ \citen{Endo00e}, shows excellent agreement with the calculated trace given by eq.\ (\ref{magres1}), where $\mu_\mathrm{W}$ and $V_1$ are used as fitting parameters. The agreement is especially good at 4.2 K, where the experimental and calculated traces are indistinguishable in Fig.\ \ref{rawtherm}(b). At 1.4 K, deviation can be observed for $B$ above $\sim$0.13 T, mainly attributable to the disturbance by the SdH effect. It is tempting to take the good agreement as evidence for the modulation being actually simply sinusoidal. We will show below, however, that the agreement does not necessarily eliminate the possibility of the presence of higher harmonics in the modulation profile.

Equation (\ref{magres1}) can readily be extended for potential modulation including higher harmonics,
\begin{equation}
V(x)=\sum\limits_{n=1}^\infty{V_n}\cos\left(\frac{2\pi}{a}nx\right),
\label{potential}
\end{equation}
since each harmonics contribute independently \cite{Gerhardts92}. The expression of CO then becomes
\begin{equation}
\frac{\Delta\rho_{xx}^\mathrm{osc}}{\rho_0}=\gamma A\left(\frac{\pi}{\mu_\mathrm{W}B}\right) |B|\sum\limits_{n=1}^\infty A\left(\frac{T}{T_{a,n}}\right) \frac{nV_n^2}{a}\sin\left(2\pi\frac{2R_\mathrm{c} }{a/n}\right).
\label{magres}
\end{equation}
The characteristic temperature for the $n$-th harmonic reduces to $1/n$ of that of the fundamental component as
\begin{equation}
T_{a,n}=\frac{1}{2\pi^2}\frac{(a/n)k_\mathrm{F}}{2}\frac{\hbar\omega_\mathrm{c}}{k_\mathrm{B}}=\frac{\tau_{a}}{n}B,
\label{chrtmp}
\end{equation}
where $\tau_{a}$$\equiv$$(1/2\pi^2)(e\hbar/2m^*)(ak_\mathrm{F}/k_\mathrm{B})$. The reduction in the characteristic temperature deeply influences the way the harmonic reveals itself in CO\@. The factor $A(T/T_{a,n})$ decreases with decreasing $B$ from unity, the asymptotic value at the high-$B$ limit, to zero; the decrease is more rapid for higher $T$, or for smaller $\tau_{a}/n$ which reflects either the decrease in $a$ or the increase in $n$. As illustrated in Fig.\ \ref{rawtherm}(c), the degree of damping is quite sensitive to the change in $T$ and $n$ in the magnetic-field and the temperature range of the present interest. As a result, higher harmonics are much more heavily damped than the fundamental component; $A(T/T_{a,n})$ works as a ``low-pass filter''. This makes it more difficult for the CO with larger $n$ to be observed in the magnetoresistance traces, especially at the higher temperature.

\begin{figure}[tb]
\includegraphics[bbllx=20,bblly=70,bburx=580,bbury=830,width=8.5cm]{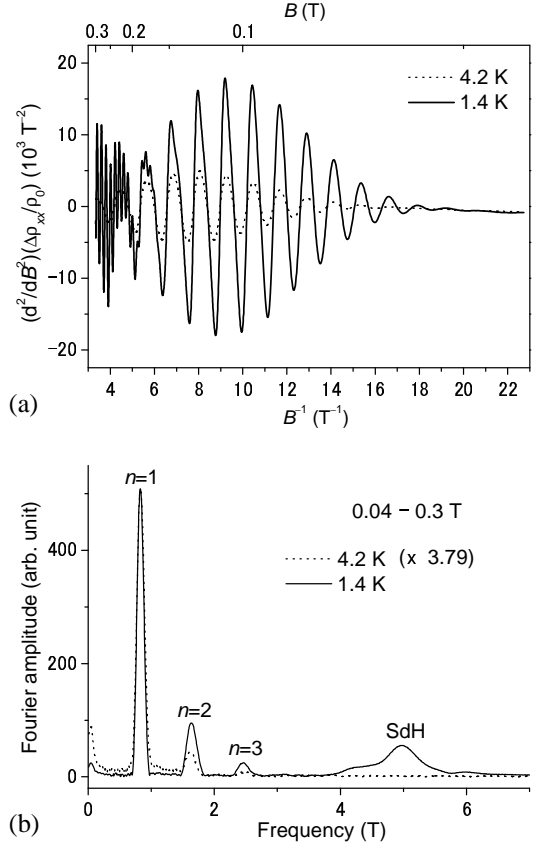}
\caption{(a) The second derivatives $(\mathrm{d}^2/\mathrm{d}B^2)(\Delta \rho_{xx}/\rho_0)$ of magnetoresistance traces shown in Fig.\ \ref{rawtherm}(a), plotted against inverse magnetic field (within the window show by the horizontal marker in Fig.\ \ref{rawtherm}(a)). (b) Fourier spectra of the traces in (a). The spectrum for 4.2 K is multiplied by a factor 3.79 to align the $n$=1 peak height.}
\label{d2dB2}
\end{figure}

The presence of higher harmonics, despite the seeming absence in Fig.\ \ref{rawtherm}(b), can be testified by Fourier spectra shown in Fig.\ \ref{d2dB2}(b). For the Fourier transform to be effective, it is advantageous to extract in advance oscillatory part from the raw magnetoresistance data. In Fig.\ \ref{d2dB2}, we made use of second derivative with respect to $B$ as a convenient way to separate out the oscillatory part. The resulting $(\mathrm{d}^2/\mathrm{d}B^2)(\Delta\rho_{xx}/\rho_0)$ is plotted against inverse field in Fig.\ \ref{d2dB2}(a) for both 1.4 and 4.2 K, whose Fourier transforms are the spectra shown in Fig.\ \ref{d2dB2}(b). The spectra exhibit peaks from the fundamental and the second and third harmonic CO as well as a peak from SdH\@. The spectra are normalized so as to equalize the height of the $n$=1 peak for both temperatures; it can readily be observed that the peak height for $T$=4.2 K relative to that of $T$=1.4 K becomes progressively smaller for $n$=2 and 3 CO, reflecting the enhanced temperature dependence of $A(T/T_{a,n})$ for increased $n$ (smaller $T_{a,n}$). The SdH peak is not resolved for $T$=4.2 K for the magnetic-field window used in the present analysis, because of its still higher sensitivity to the temperature.

\begin{figure}[tb]
\includegraphics[bbllx=20,bblly=70,bburx=580,bbury=830,width=8.5cm]{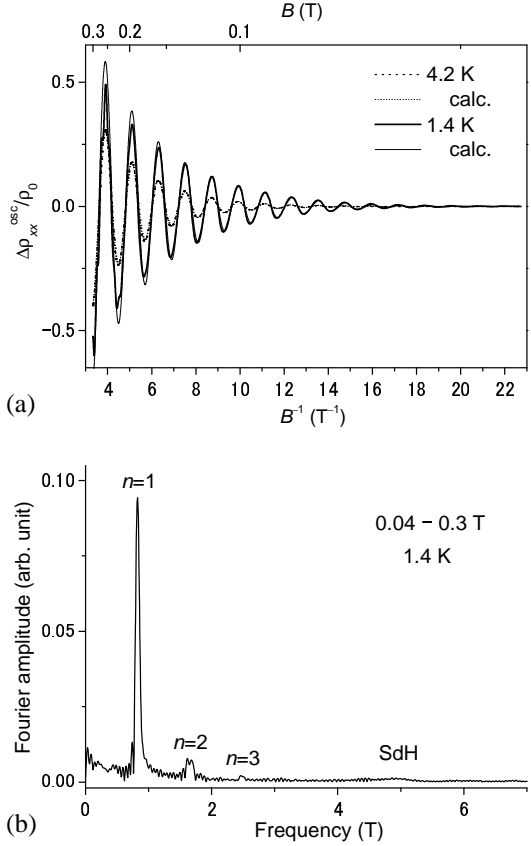}
\caption{Oscillatory parts of the magnetoresistance (replot of Fig.\ \ref{rawtherm}(b)), plotted against inverse magnetic field (within the window show by the horizontal marker in Fig.\ \ref{rawtherm}(a)). Calculated curves using eq.\ (\ref{magres1}) are also plotted by thin solid (1.4 K) and dotted (4.2 K) lines. (b) Fourier spectrum of the 1.4 K trace in (a) after divided by the $B$-dependent prefactor $F_1(B)$.}
\label{oscpart}
\end{figure}

It should be cautioned that the second derivative works as a sort of high-pass filter that emphasizes rapidly varying component, the higher harmonics in the present context. This remains innocuous for qualitative argument of, say, the temperature dependence of a particular peak. For quantitative comparison between different peaks, however, this approach is obviously inadequate. Instead, raw magnetoresistance after subtracting the slowly-varying background as was done in Fig.\ \ref{rawtherm}(b) should be analyzed directly. According to eq.\ (\ref{magres}), $\Delta\rho_{xx}^\mathrm{osc}/\rho_0$ includes the $n$-th harmonic each weighted by the $B$- and $n$-dependent envelope function $F_n(B)$$\equiv$$A(\pi/\mu_\mathrm{W}B)$$A(T/T_{a,n})$$|B|$. Therefore, a value proportional to $nV_n^2$ is obtained by performing Fourier transform to $\Delta\rho_{xx}^\mathrm{osc}/\rho_0$ divided by $F_n(B)$, plotted as a function of $B^{-1}$. An important point to keep in mind is the necessity to use different $F_n(B)$ depending on $n$ to deduce $nV_n^2$ from the same $\Delta\rho_{xx}^\mathrm{osc}/\rho_0$. Figure \ref{oscpart}(b) shows a Fourier spectrum taken from $\Delta\rho_{xx}^\mathrm{osc}/\rho_0$, shown in Fig.\ \ref{oscpart}(a), after divided by $F_1(B)$. The peak at 0.82 T is proportional to $1\times V_1^2$. Evaluation of $2V_2^2$ and $3V_3^2$, however, requires Fourier spectra different from that shown in Fig.\ \ref{oscpart}(b), namely, the ones obtained from the oscillatory part divided by $F_2(B)$ and $F_3(B)$, respectively, instead of $F_1(B)$. Repeating the Fourier transforms, we can get the values proportional to $2V_2^2$ and $3V_3^2$, hence the ratio $|V_2/V_1|$ and $|V_3/V_1|$, from the heights of the peaks at 1.64 and 2.46 T of the appropriate Fourier spectra. Note that only the absolute value of $V_n$ can be deduced from CO, since the amplitude of the CO depends only on the square of $V_n$'s [see eqs.\ (\ref{magres1}) and (\ref{magres})] and their signs are irrelevant.

There is another, more practical way to evaluate $n$=2 and 3 harmonic components from a single Fourier spectrum such as shown in Fig.\ \ref{oscpart}(b); the function to be Fourier transformed can be rewritten as that for $n$=1 multiplied by $F_1(B)/F_n(B)$=$A(T/T_{a,1})$/ $A(T/T_{a,n})$, and since the correction factor is much more slowly varying function of $b\equiv B^{-1}$ than the rest of the function including CO, it can be taken out of the integral of the Fourier transform and replaced by its average over the window $[B_f^{-1},B_i^{-1}]$$\equiv$$[b_f,b_i]$ for the transformation. The corrected peak heights are thus obtained by multiplying
\begin{equation}
\left\langle \frac{A(T/T_{a,1})}{A(T/T_{a,n})} \right\rangle=\frac{1}{b_i-b_f}\int_{b_f}^{b_i}\frac{A(Tb/\tau_{a})}{A(nTb/\tau_{a})}\mathrm{d}b
\label{aveA}
\end{equation}
to the corresponding peaks in Fig.\ \ref{oscpart}(b). Equation (\ref{aveA}) can readily be calculated analytically for $n$=2 and 3 \cite{aveA23}.

In order to determine harmonic contents $|V_n|$ ($n$=1, 2, 3) as accurately as possible from CO, we take the following strategy. First, we obtain $|V_1|$ by fitting eq.\ (\ref{magres1}) to $\Delta\rho_{xx}^\mathrm{osc}/\rho_0$ at 4.2 K\@. The higher temperature is selected because higher harmonics are more effectively damped (see Fig.\ \ref{rawtherm}(c)) and therefore their contribution to CO can be more safely neglected. Second, the ratio $|V_2/V_1|$ and $|V_3/V_1|$ are deduced from Fourier spectra using $\Delta\rho_{xx}^\mathrm{osc}/\rho_0$ at 1.4 K, since sensitivity to higher harmonics increases at lower temperature. The analysis is done for $\Delta\rho_{xx}^\mathrm{osc}/\rho_0$ normalized by adequate envelope function for $n$, $F_n(B)$, or employing $F_1(B)$ and then using the correction factor eq.\ (\ref{aveA}), both giving consistent results. Finally, $|V_2|$ and $|V_3|$ are calculated from the value of $|V_1|$ and the ratios $|V_2/V_1|$ and $|V_3/V_1|$. Here we rely on our assumption that the harmonic contents do not vary with temperature, so long as the temperatures are already low enough.

\begin{figure}[tb]
\includegraphics[bbllx=20,bblly=70,bburx=580,bbury=830,width=8.5cm]{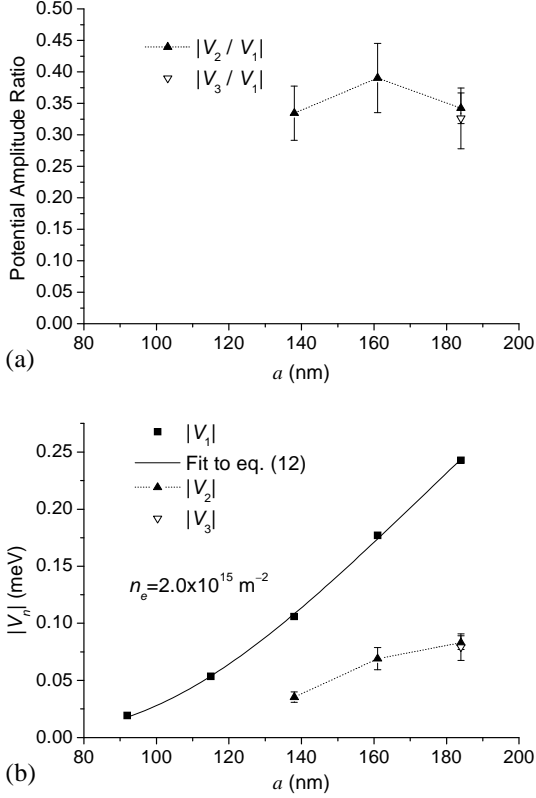}
\caption{(a) The ratio of the amplitude of the 2nd and 3rd harmonic potential to that of the fundamental component, obtained by the analysis of the Fourier spectra detailed in the text, plotted against the period of ULSL\@. (b) The amplitude of the potential modulation versus the period. The fundamental component $V_1$ was obtained by fitting the experimental 4.2 K traces to eq.\ (\ref{magres1}). Harmonics $V_2$ and $V_3$ were deduced using $V_1$ and the ratio plotted in (a). $V_1$ vs $a$ is fitted to eq.\ (\ref{expdecay1}) with $w$=4.8 nm, and using $\phi_1$ and $d$ as fitting parameters; excellent fitting was attained with $|\phi_1|$=72.3 mV and $d$=97.7 nm. Dotted line for $V_2$ is just an eye-guide.}
\label{adep}
\end{figure}

The ratios and $|V_n|$'s thus obtained for five ULSL samples are plotted in Figs.\ \ref{adep}(a) and \ref{adep}(b), respectively, against the period $a$. The second harmonic is observed for ULSL with $a$$\geq$138 nm, and the third harmonic only for $a$=184 nm. Note however that the absence of the corresponding peaks in the Fourier spectra for smaller period ULSL does not necessarily attest to the nonexistence of the higher harmonics. It simply reveals that they are less than the limit of their observation through CO; the limit becomes severer for smaller $a$ because of the dependence of the thermal damping factor on $a$, as has already been discussed thoroughly above.

The ratios $|V_2/V_1|$ and $|V_3/V_1|$ plotted in Fig.\ \ref{adep}(a) were acquired from the average of typically ten data points of the analyses described above, repeated for values of $n_\mathrm{e}$ ranging from 1.9 to 3.1$\times$10$^{15}$ m$^{-2}$ varied by LED illumination. We expected, and actually observed, that the ratios do not show systematic dependence on $n_\mathrm{e}$. The averaging was done to reduce statistical scattering that could have crept in during the measurement and/or the analysis, and the error bars represent the standard deviation. The ratios are roughly 0.3 and do not depend much on $a$. Figure \ref{adep}(b) illustrates that $|V_1|$ rapidly decrease with decreasing $a$. The plot is for $n_\mathrm{e}$=2.0$\times$10$^{15}$ m$^{-2}$, and $|V_1|$ showed slight decrease with increasing $n_\mathrm{e}$ as has been detailed in ref.\ \citen{Endo05m}. The dependence of $|V_n|$ on $a$ will be discussed in the subsequent section. It is easy to show that the oscillatory part reconstructed using eq.\ (\ref{magres}), substituting the values of $|V_n|$'s obtained here, introduces only negligibly small modification to that calculated by eq.\ (\ref{magres1}) using $|V_1|$ alone, owing to the $V_n^2$ dependence as well as to the factor $A(T/T_{a,n})$, in full support of the properness of the prescription we have followed to deduce $|V_n|$'s.

\section{Discussion}
\label{Discussion}
\subsection{The grating-induced effect localized at the surface}
\label{locsurf}
We start by a simplistic model that (i) any alteration in the GaAs/AlGaAs crystal introduced directly by our grating is localized at the surface, and (ii) the charge that might be present between the surface and the heterointerface is negligibly small. Then the electrostatic potential $\phi(x,z)$ inside the crystal is given by the solution of the Laplace's equation $\nabla^2$$\phi(x,z)$=0, with the boundary conditions imposed by the grating at the surface, 
\begin{equation}
\phi(x,0)=\sum\limits_{n=0}^\infty\phi_n \cos\left( \frac{2\pi}{a}nx\right),
\label{surface}
\end{equation}
and deep in the substrate, $\phi(x,z)\rightarrow 0$ ($z\rightarrow\infty$). Here, we have taken the $z$-axis normal to the 2DEG plane pointing inward the crystal, with $z$=0 denoting the surface. The harmonic function that satisfies the boundary conditions is simply given by
\begin{equation}
\phi(x,z)=\sum\limits_{n=0}^\infty\phi_n \exp\left(-\frac{2\pi}{a}nz\right) \cos\left(\frac{2\pi}{a}nx\right),
\label{atz}
\end{equation}
with $\phi_n$'s defined in eq.\ (\ref{surface}). The potential modulation seen by the electrons at the 2DEG plane at $z$=$d$, eq.\ (\ref{potential}) \cite{omitcnst}, is given after taking into account the screening by the 2DEG electrons themselves. Using Thomas-Fermi approximation for the screening, the Fourier components become
\begin{equation}
V_n=(-e)\phi_n \exp\left(-\frac{2\pi}{a}nd\right)\left/ \epsilon_\mathrm{TF}\left(\frac{2\pi}{a}n\right) \right.,
\label{expdecay}
\end{equation}
with
\begin{equation}
\epsilon_\mathrm{TF}(q)=1+\frac{2}{a_\mathrm{B}^*q}F(q),
\label{epsilon}
\end{equation}
where $a_\mathrm{B}^*$=$4\pi\epsilon_0\epsilon\hbar^2/m^*e^2$ represents the effective Bohr radius (10.4 nm for GaAs with the relative dielectric constant $\epsilon$=13.18 and the effective mass $m^*$=0.067$m_\mathrm{e}$) and $F(q)$ the form factor reflecting the finite thickness of the electron wave function \cite{Price84,Hirakawa86,Esfarjani90,Endo05m}. The Thomas-Fermi approximation is appropriate for the length scale longer than half the Fermi wavelength \cite{AndoR82,Hirakawa86,Endo05m}, which is actually the case for the range of $n_\mathrm{e}$ encompassed in the present study, even for our shortest length scale 184/3 nm. Note that $\epsilon_\mathrm{TF}(q)$ decreases with increasing $q$ \cite{FormFactor}, so that smaller length scale (higher harmonics) survives the screening better. The form factor may be evaluated numerically using numerically calculated wave function \cite{Endo05m}. Instead, we employ in the present study $F(q)$ calculated using Fang-Howard wave function \cite{Fang66} $\zeta^\mathrm{FH}(z)$=$\Theta(Z)\sqrt{b^3/2}Z\exp(-bZ/2)$ with $Z$$\equiv$$z-z_0$ and $z_0$ representing the location of the heterointerface ($z_0$=90 nm for the present samples) and $\Theta(Z)$ the unit step function,
\begin{equation}
F^\mathrm{FH}(q,b)=\frac{1+(9/8)(q/b)+(3/8)(q/b)^2}{(1+q/b)^3},
\label{formFH}
\end{equation}
and ${\epsilon_\mathrm{TF}}^\mathrm{FH}(q,b)$=$1+2F^\mathrm{FH}(q)/a_\mathrm{B}^* q$. The analytical formula is of benefit to the fitting to the experimental plots. The parameter $b$ is a measure of the thickness of the 2DEG wave function: $w$=$\sqrt{3}/b$ with $w$ the rms thickness. Instead of using the well-known variational formula that determines $b$ from the depletion charge and $n_\mathrm{e}$ \cite{AndoR82}, we adopt the value of $b$ that equalizes $w$ to the value deduced by numerically solving the Shr\"odinger equation and Poisson's equation self-consistently, as described in ref.\ \citen{Endo05m}. The formula for the fundamental component,
\begin{equation}
V_1=(-e)\phi_1 \exp\left(-\frac{2\pi}{a}d\right)\left/ {\epsilon_\mathrm{TF}}^\mathrm{FH}\left(\frac{2\pi}{a},\frac{\sqrt{3}}{w} \right) \right.,
\label{expdecay1}
\end{equation}
with $w$=4.8 nm (the value for $n_\mathrm{e}$=2.0$\times$10$^{15}$ m$^{-2}$) is fitted to the experimental plot using $\phi_1$ and $d$ as fitting parameters, and is also shown in Fig.\ \ref{adep}(b). The excellent agreement is achieved with the parameters $|\phi_1|$=72.3 mV and $d$=97.7 nm, the latter being quite close to the designed value of 98.7 nm. The value of $\phi_1$ is also reasonable, as will be discussed later. The good agreement suggests that the model adopted to deduce eq.\ (\ref{expdecay1}) gives a satisfactory description for the fundamental component, namely, the origin of $V_1$ can be traced back to be located at the surface. For $V_2$ and $V_3$, similar fitting is difficult because of the lack of enough number of data points. However, even a cursory glance at the plot reveals it difficult to describe $V_2$ well with eq.\ (\ref{expdecay}): according to eq.\ (\ref{expdecay}), $V_2$ should decrease more rapidly with decreasing $a$ than $V_1$, which is the opposite to the situation observed in Fig.\ \ref{adep}(b). So far, we have tacitly assumed that the $\phi_n$'s determined by the boundary condition at the surface do not depend on $a$, a reasonable assumption since we prepared different ULSL samples under the same conditions, employing the same GaAs/AlGaAs 2DEG wafer and the identical combination of EB resist and dose, hence having practically the same resist-film thickness ($\simeq$30 nm) and the elastic properties. The dependence of $V_n$ on $a$ will alter if we discard this assumption. However, it is still quite unlikely that the observed $V_2$ and $V_3$ are compatible with eq.\ (\ref{expdecay}). Their ratio to $V_1$ using eq.\ (\ref{expdecay}),
\begin{equation}
\frac{V_n}{V_1}=\frac{\phi_n\epsilon_\mathrm{TF}(2\pi/a)}{\phi_1\epsilon_\mathrm{TF}(2\pi n/a)}\exp \left[-\frac{2\pi}{a}(n-1)d\right],
\label{potratio}
\end{equation}
becomes very small because of the exponential factor, unless the ratio $\phi_n/\phi_1$ is unacceptably large. For example, $|\phi_2/\phi_1|$$\simeq$5 and $|\phi_3/\phi_1|$$\simeq$100 is required to explain the experimental $|V_2/V_1|$ and $|V_3/V_1|$ by eq.\ (\ref{potratio}) for $a$=184 nm, and the values are still larger for smaller $a$. This implies that the origin of the second and the third harmonics must be sought after in the grating-induced effect beyond the effect localized at the surface.

\subsection{Comparison with the theories of strain-induced potential modulation} The strain induced by the grating is, of course, not perfectly localized at the surface but propagate inside the crystal. Therefore the strain can in principle generate potential modulation at an arbitrary depth $z$ through deformation potential or piezoelectric effect. Analytic expressions are presented in the theories of strain-induced potential modulation \cite{Davies94,Larkin97}. Here we quickly review the results of the theories essential to the present discussion. Writing Airy's stress function as $z\chi$ with $\chi$ a harmonic function \cite{notation}, $\chi(x,z)=\sum\nolimits_{n=0}^\infty\chi_n \exp\left(-q_nz\right) \cos\left(q_nx\right)$, with $\chi_n$'s determined by \textit{elastic} boundary condition (compatibility with the grating) at the surface $\chi(x,0)$ and $q_n$$\equiv$$2\pi n/a$, Fourier coefficients for the deformation potential is given by,
\begin{equation}
\psi_n^\mathrm{def}(z)=-\frac{\Xi}{(-e)}\frac{2(1+\nu)(1-2\nu)}{E}q_n\chi_n\exp\left(-q_nz\right),
\label{deformation}
\end{equation}
where $\Xi$, $\nu$, and $E$ represent the deformation potential constant, Poisson's ratio, and Young's modulus of the host semiconductor crystal, respectively. The piezoelectric charge density $\rho(x,z)$ for the angle $\theta$=45$^\circ$ between the [100] axis and the $x$ axis reads in Fourier coefficients
\begin{equation}
\rho_n(z)=\frac{1}{2}d_{14}\left[(5-2\nu)-3q_nz\right]q_n^2\chi_n\exp\left(-q_nz\right),
\label{piezocharge}
\end{equation}
with $d_{14}$ the only surviving element of the piezoelectric tensor in the crystal with $F\bar{4}3m$ symmetry. The piezoelectric potential is found by solving Poisson's equation $\nabla^2$$\psi^\mathrm{pz}(x,z)$=$-\rho(x,z)/\epsilon_0\varepsilon$. It can generally be written as
\begin{equation}
\psi_n^\mathrm{pz}(z)=\frac{d_{14}}{8\epsilon_0\epsilon}\left[C_n+(7-4\nu)q_nz-3(q_nz)^2\right]\chi_n\exp\left(-q_nz\right),
\label{piezo}
\end{equation}
with $C_n$ constants to be determined by boundary conditions. The condition $\psi^\mathrm{pz}(x,z)\rightarrow 0$ ($z\rightarrow\infty$) is automatically satisfied. In ref.\ \citen{Larkin97}, an equipotential layer at $z$=$c$ (either the surface or the $\delta$-doped layer), $\psi^\mathrm{pz}(x,c)$=0, is considered, which results in $C_n$=$3(q_nc)^2-(7-4\nu)q_nc$. For electrons at 2DEG, $V_n^\mathrm{def}$=$(-e)\psi_n^\mathrm{def}(d)/\epsilon_\mathrm{TF}(q_n)$ and $V_n^\mathrm{pz}$=$(-e)\psi_n^\mathrm{pz}(d)/\epsilon_\mathrm{TF}(q_n)$ after screened by the 2DEG \cite{epsTF}. Note that both $\psi_n^\mathrm{def}(z)$ and $\psi_n^\mathrm{pz}(z)$, as well as $\phi_n\exp(-q_nz)$ in the previous section, include the factor $\exp(-q_nz)$ and therefore attenuate with $z$. The attenuation is more rapid for smaller $a$ or for higher harmonics. What is peculiar in $\psi_n^\mathrm{pz}(z)$ is its inclusion of the polynomial of $q_nz$, leading to non-monotonic behavior with $z$ around $z$$\sim$$a/n$. Therefore, $V_n^\mathrm{pz}/V_1^\mathrm{pz}$ can become large when $z$=$d$ happens to lie near the node of $\psi_1^\mathrm{pz}(z)$ without resorting to unnaturally large $\chi_n/\chi_1$.

\begin{figure}
\includegraphics[bbllx=20,bblly=20,bburx=580,bbury=750,width=8.5cm]{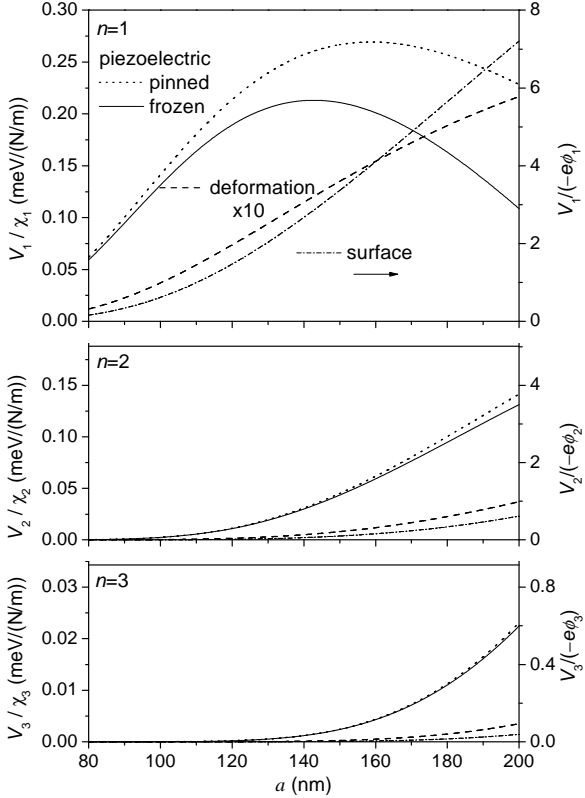}
\caption{Normalized piezoelectric and deformation potentials (left axis), $V_n^\mathrm{pz}/\chi_n$ and $V_n^\mathrm{def}/\chi_n$ ($n$=1, 2, 3), as a function of $a$, calculated with eqs.\ (\ref{piezo}) and (\ref{deformation}), respectively, and  $\epsilon_\mathrm{TF}(q_n)$=${\epsilon_\mathrm{TF}}^\mathrm{FH}(q_n,\sqrt{3}/w)$, for our sample parameters $d$=98.7 nm, $w$=4.8 nm. Elastic parameters for GaAs \cite{Larkin97,Adachi85}, $\Xi$=$-$8.2 eV, $\nu$=0.31, $E$=85.3 GPa, and $d_{14}$=$-$2.69$\times$10$^{-12}$ mV$^{-1}$ are used. For $V_n^\mathrm{pz}/\chi_n$, two different boundary conditions at the surface, ``pinned'' and ``frozen'' surface models, are considered and are plotted by dotted and solid lines, respectively. $V_n^\mathrm{def}/\chi_n$ (dashed lines) are magnified by a factor 10. $V_n/(-e\phi_n)$ calculated by eq.\ (\ref{expdecay}) with $\epsilon_\mathrm{TF}(q_n)$=${\epsilon_\mathrm{TF}}^\mathrm{FH}(q_n,\sqrt{3}/w)$ are plotted by dash-dotted line (right axis) for comparison. Note the expanded ($\times$5) vertical scale (both right and left axes) for $n$=3.}
\label{calcadep}
\end{figure}

Unfortunately, it is not possible to make a quantitative comparison between our data and the theories, since the evaluation of $\chi_n$'s requires the knowledge of the elastic constants for our EB resist, which are currently unavailable \cite{metal}. For qualitative comparison of the dependence on $a$, we plot $V_n^\mathrm{def}$/$\chi_n$ and $V_n^\mathrm{pz}$/$\chi_n$ ($n$=1, 2, 3) against $a$ in Fig.\ \ref{calcadep}, with $\epsilon_\mathrm{TF}(q_n)$=${\epsilon_\mathrm{TF}}^\mathrm{FH}(q_n,\sqrt{3}/w)$ and using elastic parameters for GaAs \cite{Larkin97,Adachi85} and structure parameters for our present samples. We also plot, with a separate (right) vertical axis, $V_1/(-e\phi_1)$ calculated by eq.\ (\ref{expdecay1}), which describes the experimental behavior well, and its equivalent for $n$=2 and 3. For the calculation of $V_n^\mathrm{pz}$, we adopted two different models for boundary conditions at the surface after refs.\ \citen{Davies94} and \citen{Larkin97}. In the ``pinned'' surface model, appropriate if the sample is in full equilibrium, we assume the surface $z$=$c$=0 to be the equipotential plane $\psi_n^\mathrm{pz}(x,c)$=0 \cite{unidope}. The ``frozen'' surface model assumes that charges at the surface state are frozen because of slow equilibration rate at low temperatures between the surface state and the 2DEG. Capacitance spectroscopy \cite{Ali99} and $n_\mathrm{e}$ vs back-gate voltage measurements \cite{Kawarazuka01} suggest that the frozen surface model is actually a better description at cryogenic temperatures. Appealing to the high dielectric constant of the semiconductor, we may take $\partial \psi_n^\mathrm{pz}(x,z)/\partial z$=0 at $z$=0 in this model, leading to $C_n$=$7-4\nu$ in eq.\ (\ref{piezo}). In contrast to the situation considered in ref.\ \citen{Davies94}, we take $\partial \psi_n^\mathrm{pz}(x,z)/\partial z$=0 not only at open surface between the resist gates but also under the gates, because our grating is made of non-conducting material; we do not need to solve mixed boundary value problem. For both boundary conditions, $V_1^\mathrm{pz}$ shows a maximum in the range of $a$ investigated in the present study, at variance with the experimental result. The dependence of $V_1^\mathrm{def}$ on $a$ appears to be closer to that of the experiment. However, as mentioned earlier and is well established \cite{Skuras97}, the strain-induced effect is dominated by the piezoelectric effect, $V_n^\mathrm{pz}$$\gg$$V_n^\mathrm{def}$, for the present crystallographic direction, which is actually observed in Fig.\ \ref{calcadep}. Therefore, neither $V_1^\mathrm{pz}$ nor $V_1^\mathrm{def}$ gives better account of the experimental $V_1$, compared with the success of the model described in \S \ref{locsurf}. An attempt to explain this will be presented in the next section. For $n$=2 and 3, our experimental data are neither sufficient in quantity nor accurate enough for detailed comparison with the theory to be meaningful. Nonetheless, Fig.\ \ref{calcadep} does suggest that the piezoelectric effect is the most likely origin of the experimentally observed large $|V_n/V_1|$ ($n$=2, 3).

\subsection{Plausible model for our potential modulation}
In the theories \cite{Davies94,Larkin97}, several simplifications are made to enable analytical treatments. Above all, elastic properties of the layers between the surface and the heterointerface are all approximated by those of a single material, GaAs, and are treated as isotropic. We suspect this simplification to be partly responsible for the disagreement between the behavior of our data and eq.\ (\ref{piezo}). In reality, 80 nm out of the 90 nm between the surface and the heterointerface is AlGaAs, with the upper half of the AlGaAs layer doped with Si. Slightly different elastic, thermal, or electrostatic properties of each layer stacked successively can in principle alter the potential introduced to the 2DEG through the layers. Specifically, the lattice mismatch between GaAs and AlGaAs is about 0.1\%, roughly the same order of magnitude as the estimated strain introduced by the grating via differential contraction \cite{Davies94,Larkin97}. Therefore the stress introduced to the GaAs cap layer at the surface from the resist is possibly modified when transmitting across the interface, 10 nm below the surface, to the Si-doped AlGaAs layer; depending on the sign of the strain at the surface, the stress will be either enhanced or reduced at the interface. If the sign and the magnitude of the strain is such that the stress is nullified at the interface, the piezoelectric charge and potential may taken to be generated only in the vicinity of the surface, realizing the situation assumed in \S \ref{locsurf}. We speculate this is basically what is taking place for the fundamental ($n$=1) component. We take here a further step to see whether the value of $\phi_1$ obtained from the experimental data in \S \ref{locsurf} is reasonably explained within this model by roughly estimating the size of the differential contraction required to obtain the value.

For simplicity, we assume the stress hence the piezoelectric charge to be generated only \textit{at} the surface, neglecting the finite thickness of the cap layer along whose depth the stress is expected to attenuate. We thus use eq.\ (\ref{piezo}) with $z$=0 for the piezoelectric potential that serves as a boundary condition eq.\ (\ref{surface}) for the electrostatic problem, resulting in 
\begin{equation}
\phi_n=\frac{d_{14}}{8\epsilon_0\epsilon} C_n\chi_n.
\label{phichi}
\end{equation}
The relation of $\chi(x,0)$ with the differential contraction $\alpha_\mathrm{gate}\Delta T$ ($\alpha_\mathrm{gate}$ denoting the difference of the linear expansion coefficient between the gate and the semiconductor) is given by the compatibility equation (the boundary condition for the elastic problem at the semiconductor surface). In an approximation named ``elastic'' gate model in ref.\ \citen{Larkin97}, the compatibility equation reads
\begin{equation}
\chi(x,0)=\frac{hE_\mathrm{gate}\alpha_\mathrm{gate}\Delta T}{1-\nu_\mathrm{gate}}u(x),
\label{compatible}
\end{equation}
where $h$, $\nu_\mathrm{gate}$, and $E_\mathrm{gate}$ represent the thickness, Poisson's ratio, and Young's modulus for the gate, and $u(x)$=1 for $x$ under the gates and 0 for $x$ in between. The ``elastic'' gate model applies to thin gates. There the gate is forced to contract with the same ratio as the underlying semiconductor, uniformly stressed; the gate reacts back a force, concentrated at the edge, on the semiconductor. The model appears to be suitable for our thin ($h$$\simeq$30 nm), and probably soft, resist gate. For a grating with duty ratio $\alpha$ ($\sim$1/2 in the present samples), $u(x)=\alpha+(2/\pi)\sum\nolimits_{n=1}^\infty [\sin(\alpha \pi n)/n] \cos\left(q_nx\right)$, thus eq.\ (\ref{compatible}) becomes in Fourier coefficient
\begin{equation}
\chi_n=\frac{hE_\mathrm{gate}\alpha_\mathrm{gate}\Delta T}{1-\nu_\mathrm{gate}}\frac{\sin(\alpha \pi n)}{\pi n/2}
\label{compatibleF}
\end{equation}
for $n$=1, 2, 3,.... From eq.\ (\ref{phichi}) and using the ``frozen'' surface model, $C_n$=$7-4\nu$, $|\chi_1|$=4.4 N/m is obtained from $|\phi_1|$=72.3 mV\@. Equation (\ref{compatibleF}) shows that the value of $\chi_1$ can be achieved with $\alpha_\mathrm{gate}\Delta T$ considerably smaller than 1\% by using reasonable values for elastic constants for the resist (\textit{e.g.}, $\nu_\mathrm{gate}$$\sim$0.3 and $E_\mathrm{gate}$ tens of GPa), although their exact values are unavailable.

Of course it is highly unlikely that the stress introduced at the surface completely relaxes at the interface between the cap and the Si-doped layer. The stress and therefore the piezoelectric charge probably remain below the interface, although smaller than predicted by eq.\ (\ref{piezocharge}). The discussion we have presented so far suggests that for the fundamental ($n$=1) component the residual piezoelectric charge can be taken to be negligibly small. For higher harmonics, the remnant piezoelectric charge takes on more important role, since the strong piezoelectric potential limited to the vicinity of the surface becomes less influential at the 2DEG plane due to the exponential factor, as discussed in \S \ref{locsurf}. We presume that the residual piezoelectric charge is mainly responsible for the higher-harmonic potential modulation seen by the 2DEG electrons. An alternative candidate for the origin of the higher harmonics are the charge building up at the heterointerface just above the 2DEG due to the difference in the elastic, piezoelectric, or dielectric constants, as discussed in ref.\ \citen{Larkin97}. It seems quite difficult, at this stage, to further elaborate and specify the origin from our currently available data.

\section{Conclusions}
We have investigated the harmonic content included in strain-induced potential modulation, with recourse to detailed analysis of CO in ULSL's. We have pointed out that the magnitude of higher harmonics can be larger than seemingly suggested by their appearance in raw CO traces, since the CO is heavily weighted in favor of slower oscillation due to the thermal damping factor $A(T/T_{a,n})$. In our present samples, $|V_2|$ and $|V_3|$ are as large as $\sim$0.3$|V_1|$, although the second or third harmonics are not readily discernible in the raw magnetoresistance data.

Higher harmonics $V_n$ ($n$$>$1) have a general tendency to be suppressed relative to the fundamental component $V_1$ owing to the ubiquitous exponential factor, $\exp(-q_nz)$, common to all the mechanisms discussed in \S \ref{Discussion}. However, for the piezoelectric effect, the dominant effect in the present samples, the dependence of $V_n$ on $q_nz$ becomes complicated because of the extra polynomial dependence, which we take as one of the causes of rather large ratio $V_n/V_1$ ($n$=2,3). The dependence of $V_n$ on $a$ is naturally explained by assuming that the source of $V_1$ is localized at the surface with negligible charge between the surface and the 2DEG, while $V_2$ and $V_3$ originate from piezoelectric (or interface) charge located closer to 2DEG\@. Noteworthily, the screening by 2DEG electrons favors higher harmonics, giving advantage of the factor of roughly 2 and 3 for $V_2$ and $V_3$, respectively.

We want to stress that the higher harmonics are shown to be able to introduce into the 2DEG plane the length scale substantially smaller than its depth $d$$\simeq$99 nm from the surface, the smallest of those detected by CO in the present experiments being $a/2$=69 nm and $a/3$=61 nm, respectively. This suggests a possibility of realizing ULSL with the period smaller than $d$ by exploiting the strain-induced piezoelectric effect, using metallic grating; since the modulation introduced electrostatically by applying a bias to the metallic grating is almost exclusively composed of the fundamental component owing to the factor $\exp(-q_nz)$, it is possible to tune the bias to cancel out the fundamental component due to the strain-induced effect, thereby making the higher harmonics selectively survive \cite{a500}. Lateral superlattice samples with small period, close to or shorter than, \textit{e.g.}, the Fermi wave length, made from a minimally disordered 2DEG residing at large depth in the MODFET structure, will prove invaluable to experimental studies of many aspects of superlattices.

\section*{Acknowledgment}
This work was supported by Grant-in-Aid for Scientific Research (C) (15540305) and (A) (13304025) and Grant-in-Aid for COE Research (12CE2004) from Ministry of Education, Culture, Sports, Science and Technology.


\begin{thebibliography}{99}
\bibitem{separate} The data for the two temperatures are acquired by separate cooling-downs (using different cryostats) and therefore the electron densities are not exactly the same. This explains small discrepancy between $A(1.4\; \mathrm{K}/T_{a,3})$ and $A(4.2\; \mathrm{K}/T_{a,1})$, which should otherwise exactly overlap each other.
\bibitem{omitcnst}
The constant term $V_0$ is omitted, which can be done without the loss of generality.
\bibitem{notation}
$\chi$ is denoted as $\phi$ in the original papers \cite{Davies94,Larkin97}. We have changed the notation so as not to be confused with our $\phi$ in \S \ref{locsurf}.
\bibitem{epsTF} In ref.\ \citen{Larkin97}, $\epsilon_\mathrm{TF}(q_n)$ slightly modified to include the effect of equipotential plane nearby the 2DEG is used.
\bibitem{unidope} Our 2DEG wafer does not contain Si $\delta$-doped layer that works as an equipotential layer between the surface and 2DEG plane as considered in refs.\ \citen{Davies94} and \citen{Larkin97}.
\bibitem{a500} Such cancellation was actually shown to be possible in our previous publication \cite{Kato97}, although the resulting period $a/3$=167 nm was still larger than $d$ (70 nm $+$ the thickness of 2DEG in that sample).
\end{thebibliography}
\end{document}